\documentclass[sigconf]{aamas}

\usepackage{balance} 
\usepackage{graphicx}
\usepackage{newtxmath}
\usepackage{multirow}
\usepackage{algorithmic}
\usepackage{algorithm}
\usepackage{subfig}
\usepackage{wrapfig}
\usepackage{float}

\setcopyright{ifaamas}
\acmConference[AAMAS '25]{Proc.\@ of the 24th International Conference
on Autonomous Agents and Multiagent Systems (AAMAS 2025)}{May 19 -- 23, 2025}
{Detroit, Michigan, USA}{A.~El~Fallah~Seghrouchni, Y.~Vorobeychik, S.~Das, A.~Nowe (eds.)}
\copyrightyear{2025}
\acmYear{2025}
\acmDOI{}
\acmPrice{}
\acmISBN{}

\title[AAMAS-2025 Formatting Instructions]{Tacit Learning with Adaptive Information Selection for Cooperative Multi-Agent Reinforcement Learning}

\author{Lunjun Liu}
\affiliation{
  \institution{College of Electrical and Information Engineering}
  \city{Changsha}
  \country{China}}
\email{barryyyliu@hnu.edu.cn}

\author{Weilai Jiang}
\affiliation{
  \institution{College of Electrical and Information Engineering}
  \city{Changsha}
  \country{China}}
\email{jiangweilai@hnu.edu.cn}

\author{Yaonan Wang}
\affiliation{
  \institution{College of Electrical and Information Engineering}
  \city{Changsha}
  \country{China}}
\email{yaonan@hnu.edu.cn}

\begin{abstract}
In multi-agent reinforcement learning (MARL), the centralized training with decentralized execution (CTDE) framework has gained widespread adoption due to its strong performance. However, the further development of CTDE faces two key challenges. First, agents struggle to autonomously assess the relevance of input information for cooperative tasks, impairing their decision-making abilities. Second, in communication-limited scenarios with partial observability, agents are unable to access global information, restricting their ability to collaborate effectively from a global perspective. To address these challenges, we introduce a novel cooperative MARL framework based on information selection and tacit learning. In this framework, agents gradually develop implicit coordination during training, enabling them to infer the cooperative behavior of others in a discrete space without communication, relying solely on local information. Moreover, we integrate gating and selection mechanisms, allowing agents to adaptively filter information based on environmental changes, thereby enhancing their decision-making capabilities. Experiments on popular MARL benchmarks show that our framework can be seamlessly integrated with state-of-the-art algorithms, leading to significant performance improvements.
\end{abstract}


\newcommand{\BibTeX}{\rm B\kern-.05em{\sc i\kern-.025em b}\kern-.08em\TeX}

\begin{document}


\pagestyle{fancy}
\fancyhead{}


\maketitle 


\section{Introduction}

Cooperative Multi-Agent Reinforcement Learning (MARL) has emerged as a robust framework for addressing practical challenges across various domains, including autonomous driving \cite{r43}, gaming \cite{r1}, swarm robotics \cite{r25,r26}, and smart grids \cite{r28,r29,r38}. Despite its success, learning complex cooperative strategies remains a major challenge. Firstly, neglecting the influence of other agents on the system introduces non-stationarity from the perspective of an individual, potentially leading to environmental instability. Additionally, as the number of agents increases, the observation space for joint actions expands exponentially, which may impede the learning process. To effectively address these challenges, the approach of Centralized Training and Decentralized Execution (CTDE) has been proposed and gained popularity in MARL. CTDE utilizes global information during training while achieving decentralized decision-making based on local information. It serves as the foundation for several prominent methods, including VDN \cite{r33}, MADDPG \cite{r23}, QMIX \cite{r30}, and COMA \cite{r8}.

Despite making progress, CTDE methods still face two major challenges. Firstly, agents in CTDE often rely on specific information for decision-making, and an excessive amount of information may overwhelm them. Secondly, many existing CTDE methods assume mutual independence among agents during training, only incorporating global information in the mixing network. This approach overlooks the importance of information sharing among agents \cite{r39}, thus hindering the establishment of cooperation. Therefore, addressing how agents handle and share information in MARL remains a pressing challenge.

Various explicit communication methods have been proposed to address the challenge of information sharing in multi-agent systems, including CommNet \cite{r32}, BicNet \cite{r27}, and NDQ \cite{r40}. To reduce the delays and costs associated with these communication methods, some approaches have shifted toward implicit communication frameworks, which aim to simulate communication by promoting mutual understanding among agents \cite{r21,r44}. However, in these implicit methods, agents often depend too heavily on others to evaluate information, rather than autonomously assessing its importance, effectively delegating decision-making to other agents. A potential solution to this issue is to allow agents to communicate during the training phase while relying solely on local observations during decision-making.

Constrained by cognitive limitations and individual perspectives, humans exhibit selectivity when receiving information. They process this information based on their knowledge and past experiences, selecting the most relevant details for the present moment. In collaborative settings, individuals often develop a tacit understanding through specific training, enabling them to accurately predict and comprehend their peers' intentions without explicit communication. Inspired by human information processing and cooperation patterns, we propose a novel framework called \textbf{S}elective \textbf{I}mplicit \textbf{C}ollaboration \textbf{A}lgorithm (SICA) for multi-agent systems. SICA is built upon the QMIX framework and can be extended to various methods based on CTDE paradigm. The framework comprises three key blocks: the Selection Block, the Communication Block, and the Regeneration Block. During training, the Selection Block assists agents in filtering information relevant to cooperation, which is then shared with other agents through the Communication Block to generate true information. Subsequently, the Regeneration Block utilizes local information to regenerate true information. Through iterative training, SICA gradually reduces reliance on true information, transitioning from a centralized to a decentralized framework.

Overall, SICA offers several advantages. First, SICA facilitates adaptive selection of input information, ensuring that the obtained information is highly relevant to decision-making, thereby enhancing the performance of CTDE methods. Second, SICA's exchanges do not require explicit communication, thereby avoiding issues like latency induced by communication and enabling deployment in diverse environments. Finally, SICA serves as a plug-and-play framework for agent policy networks, capable of seamlessly integrating with any value decomposition algorithm.

We evaluated the performance of SICA in the StarCraft Multi-Agent Challenge (SMAC) \cite{r31}, SMACv2 \cite{r6}, and Google Research Football (GRF) \cite{r19}. The experimental results demonstrate that SICA outperforms traditional CTDE methods and explicit communication methods in terms of performance. 


\section{Related Work}

\textbf{CTDE Framework}\quad The Centralized Training and Decentralized Execution (CTDE) framework presents a novel MARL approach aimed at addressing the scalability challenges of centralized learning \cite{r3} and the environmental non-stationarity of decentralized learning \cite{dec}. This framework operates by training agents based on global information while making decisions based on local information. Two prominent works based on this framework are VDN \cite{r33} and QMIX \cite{r30}, which introduce value decomposition methods. VDN aggregates the value functions of individual agents by directly summing them to obtain the joint action function, while QMIX further optimizes this process by ensuring the Individual-Global Maximum (IGM) condition through monotonicity. More recent methods, such as QPLEX \cite{r37}, introduce double adversarial networks to overcome representation limitations. Additionally, CTDE encompasses a series of centralized critic methods, including MADDPG \cite{r23}, COMA \cite{r8}, FOP \cite{r42},  etc., which do not impose any restrictions on the representation of the joint action-value function.

\textbf{Continous Communication}\quad CTDE methods typically incorporate global information solely in the mixing network, with agents limited to observing only local information. To augment the decision-making capabilities of agents, numerous studies have adopted communication methods to introduce global information to the agents. One pioneering work in this domain is RIAL with DIAL \cite{r7}, which aims to facilitate communication in CTDE through predefined topological structures. Meanwhile, CommNet was introduced as the first differentiable framework in MARL \cite{r32}. Subsequent works, such as ATOC \cite{r17} and I2C \cite{r5}, utilize state-dependent communication graphs to address these challenges. ATOC and I2C also introduce gating mechanisms to regulate communication links between agents. Recently, novel communication methods such as TarMAC \cite{r4} and NDQ \cite{r40} have been proposed. TarMAC employs multi-layer attention modules for multi-round communication to strengthen connections between agents. NDQ aims to reduce communication overhead by learning nearly decomposable value functions.

\textbf{Communication-Free Execution}\quad In certain scenarios, factors such as delays and costs can impede the practical deployment of communication methods. To address this challenge, some approaches permit communication only during the training phase to establish cooperative patterns among agents, while prohibiting communication during the execution phase. Previous work introduced a multi-agent framework that utilizes actions as a form of implicit communication, successfully applying it to various robotic tasks \cite{r18}. Building on this concept, researchers developed the PBL framework \cite{r35}, which fosters implicit communication through actions and introduces a secondary reward to incentivize this behavior. Furthermore, PBL incorporates a social influence reward specifically for Sequential Social Dilemma (SSD) multi-agent environments \cite{r16}, enabling agents to cooperate more effectively in SSD scenarios.

\textbf{Our Works}\quad While the aforementioned communication-free execution methods rely on gradient flow to facilitate cooperation—thereby increasing model complexity—our approach shares similarities with COLA \cite{r41} and TACO \cite{r21}. Both methods convey information through specific communication protocols and gradually transition to a decentralized framework. Additionally, to address the challenge of agent information filtering, we designed an adaptive selection mechanism, which effectively enhances the agents' information processing capabilities.


\section{Preliminary}

\textbf{Problem Formulation}\quad In a multi-agent cooperative environment, our work adheres to the definition of the Decentralized Partially Observable Markov Decision Process (Dec-POMDP) \cite{r2}, denoted as ($\mathcal{N}$, $\mathcal{S}$, $U$, $O$, $\Omega$, $P$, $r$, $\gamma$). Here, $\mathcal{N}$ represents the set of $n$ agents, and $\mathcal{S}$ is the global state space of the environment. The joint action space $\boldsymbol{u}=\{u_{1}, u_{2}, ...,  u_{n}\}\in\boldsymbol{U}=U^{n}$ consists of the independent actions of each agent, where $\gamma$ represents the discounted factor. In the Partially Observable Markov Decision Process (POMDP), agents lack access to global information. Consequently, at each time step $t$, each agent $i$ can only observe that $o_{i} \in O$ based on the observation function $\Omega(s, u)$: $\mathcal{S} \times \mathcal{U} \rightarrow O$. Agents then select an action $u_{i}$ according to the policy $\pi(u_{i}|o_{i})$. Subsequently, the state transition function $P(s^{\prime}|s, \boldsymbol{u})$ is updated to the next state $s^{\prime}$. Agents collectively receive global rewards according to a reward function $r(s, \boldsymbol{u})$ and share them. The objective for all agents is to find an optimal joint policy $\pi^{*}$ to maximize the global expected return. Our aim is to achieve this goal by designing effective implicit communication between the agents.

\begin{figure*}[t]
	\centering
	\includegraphics[scale=0.75]{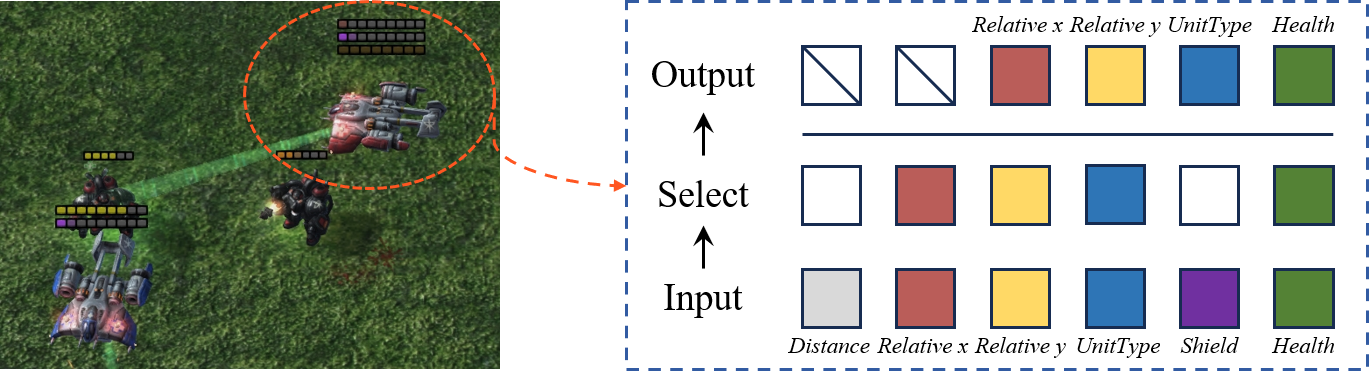}
	\caption{A case study in selection mechanism. The colored sections signify the information that the agent elects to remember, whereas the white sections denote the information that the agent chooses to ignore.}
	\label{SM}
	\vskip -0.15in
\end{figure*}

\textbf{The S6 Layer} \quad Given an input scalar $x(t)$, we consider a continuous-time invariant State Space Model (SSM) defined by the following first-order differential equation:
\begin{equation}
\begin{aligned}
\dot{h}(t)=Ah(t)_{k-1}+Bx(t)_{k}, \quad y(t)=Ch(t)_{k-1}+Dx(t)_{k}
\end{aligned}
\end{equation}
Here, $x(t)$ is an input function mapped to produce the output $y(t)$. Previous research has shown that the SSM can capture remote dependencies by initializing matrix $A$ using the HIPPO matrix \cite{r10}. Similarly, $D$ is interpreted as a parameter-based skip connection and set  to 0, following prior work \cite{r11,r14}. Additionally, as the SSM operates on continuous sequences, it is discretized using the Zero-Order Hold method, considering the discrete matrices $\bar{A}$ and $\bar{B}$:
\begin{equation}
 \begin{aligned}
 \label{matrix}
\bar{A}=\exp(\Delta A), \quad\bar{B}=(\Delta A)^{-1}(\exp(\Delta A)-I)\cdot \Delta B
 \end{aligned}
\end{equation}
From this, we can rewrite the definitional formula for SSM:
\begin{equation}
 \begin{aligned}
h_{k}=\bar{A}h_{k-1}+\bar{B}x_{k}, \quad y_{k}=Ch_{k}
 \end{aligned}
\end{equation}
At this point, the parameters $A$, $B$, $C$, $\Delta$ are static, but they are transformed into dynamic parameters to integrate a selection mechanism:  $S_{B}(x)=Linear_{N}(x)$, $S_{C}(x)=Linear_{N}(x)$, $S_{\Delta}(x)=Linear_{D}(x)$, $\tau_{\Delta}=softplus$. $Linear$ refers to linear projection. This adjustment enables the model to filter out irrelevant information and retain relevant information over time, enhancing its ability to effectively address the problem.

\textbf{Value Decomposition} \quad The introduction of value decomposition methods aims to balance the fitting capability of the Q-function with the simplicity of finding its maximum value. Currently, most value decomposition methods adhere to the Individual-Global-Max (IGM) principle \cite{QTRAN}:
\begin{equation}
\begin{split}
\mathop{\arg\max}\limits_{\boldsymbol{u}}Q_{joint}(\boldsymbol{\tau}, \boldsymbol{u})=\left( \mathop{\arg\max}\limits_{\boldsymbol{u_{1}}}Q_{1}(\tau_{1}, u_{1}), \cdots, \mathop{\arg\max}\limits_{\boldsymbol{u_{n}}}Q_{n}(\tau_{n}, u_{n}) \right)
\end{split}
\end{equation}
This implies that the joint action, which is composed of local greedy actions chosen by each agent based on their individual Q-functions, is equivalent to the joint greedy action chosen based on $Q_{joint}$. Two classic value decomposition baselines are VDN \cite{r33} and QMIX \cite{r30}. Taking QMIX as an example, the estimated value $Q_{tot}$ adopts the design of a monotonic mixing network design, while each agent has its own policy network. Both networks contain learnable parameters $\theta = \{\theta_{u}, \theta_{v}\}$, which can be updated by minimizing the Temporal Difference (TD) Loss:
\begin{equation}
 \begin{aligned}
 \label{TDLOSS}
 \mathcal{L}_{TD}(\theta)=E_{\mathcal{D}}[(y_{tot} - Q_{tot}(\boldsymbol{\tau},\boldsymbol{u},s;\theta))^{2}]
 \end{aligned}
\end{equation}
Here, $y_{tot} = r + \gamma\max\limits_{\boldsymbol{u^{\prime}}} Q^{\prime}_{tot}(\boldsymbol{\tau}^{\prime},\boldsymbol{u}^{\prime},s^{\prime};\theta^{-}))$, $E[\cdot]$ denotes the expectation function, and $\mathcal{D}$ represents the replay buffer of the transition.

\section{SICA}

In this section, we delve into the core details of SICA, outlining its adaptive information selection mechanism, communication strategy, and the transition from explicit to implicit communication. We then elucidate the overarching objectives of the agents and the learning process. Notably, our proposed method operates within the agent network and can be seamlessly integrated with any value decompostition method.

\subsection{Adaptive information selection}
In cooperative multi-agent tasks, a crucial aspect is the agents' capacity to autonomously filter input information. This involves selecting the most relevant information for ongoing cooperation, guided by experience and task requisites, while minimizing attention to irrelevant data. Currently, there is no CTDE method that equips agents with such discernment. To address this, we propose an adaptive information selection mechanism for agents. This mechanism operates by integrating historical information at each time step, giving more detail to data closer to the current time step while abstracting information that is further away. Subsequently, based on the prevailing cooperation demands, agents discern effective information and filter out irrelevant data. For instance, Figure \ref{SM} illustrates this selection mechanism in action within the SMAC environment, where two Marines and one Medivac from the red team engage a blue Medivac. Considering the red Medivac, given the proximity of both factions, it disregards distance information. Additionally, with no units generating shields on the field, the red Medivac also discounts shield information.

\subsection{SICA Framework}
\begin{figure*}[t]
\centering
\centerline{\includegraphics[scale=0.43]{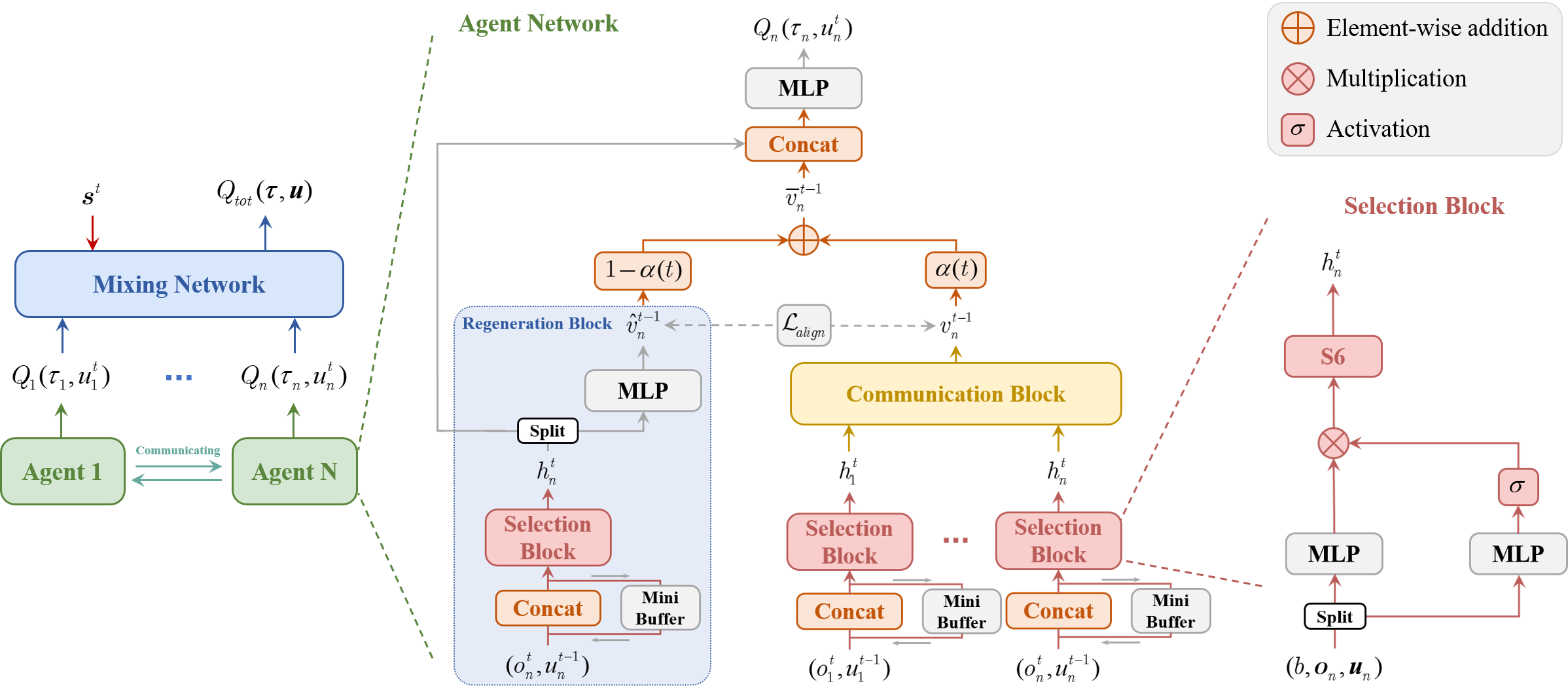}}
\caption{The overall framework of SICA. We illustrate the network architecture using the example of the $Nth$ agent. The overall architecture comprises a mixing network and agent networks ($left$). Details of the agent network include the Selection Block, Communication Block and Regeneration Block ($middle$). The Selection Block ($right$)}
\label{Framework}
\vskip -0.1in
\end{figure*}

To address the mentioned challenges and facilitate framework extension, SICA adopts the classic QMIX \cite{r30} framework. This framework comprises a mixing network and agent networks. However, unlike QMIX, SICA does not presume agent independence during training. SICA primarily enhances the agent network, as depicted in the overall framework in Figure \ref{Framework} left. This framework encompasses the Selection Block, Communication Block, and Regeneration Block, as delineated in Figure \ref{Framework} middle. The Selection Block empowers the framework with information selection capabilities. Agents can dynamically choose desired input information, as elucidated in Figure \ref{Framework} right. 

\textbf{Selection Block} \quad The Selection Block consists of two MLPs and an S6 layer \cite{r13}. Since the time intervals of the inputs in the selective tasks are variable, a time-varying model is required, so we integrated the S6 layer into the framework. The two MLPs and other components can be modeled as a Gating Unit (GU) \cite{gMLP}, which is responsible for learning long-term dependencies in the input. Therefore, the entire module can be seen as a dual selection mechanism that combines the gating mechanism with the S6 selection mechanism. Additionally, to empower the Selection Block to thoroughly select information, we establish a mini-buffer preceding it. This mini-buffer preserves the preceding $b$ observation-action pairs of a single agent, constituting a mini-batch $\{(o_{i}^{t-1},u_{i}^{t-2}), \cdots, \\(o_{i}^{t-b},u_{i}^{t-b-1})\}$ transmitted to the Selection Block. Here, $b$ represents the capacity of the mini-buffer and $i$ represents the agent $i$, both acting as hyperparameters.

Agent $i$'s observation-action pairs from the current time step and the previous $b$ time steps are integrated and passed through a GU before being fed to the S6 layer. Within the S6 layer, $\Delta$, $B$, and $C$ function as dependencies on the input, rendering the parameters data-dependent in SICA. $\Delta$ aids in controlling whether the agent prioritizes the current hidden state or the input, while $B$ and $C$ facilitate the filtering out of irrelevant information. As only the hidden state information from S6 is required, solely $B$ is utilized. The computation process for the entire module is outlined as follows:
\begin{equation}
\begin{aligned}
 \\
z_{i}^{t}&=\text{MLP}(x_{i1}^{t}) \sigma(\text{MLP}(x_{i2}^{t})) \\
h_{i}^{t}&=\bar{A}h_{i}^{t-1}+\bar{B}z_{i}^{t}
\end{aligned}
\end{equation}
Here, $\{x_{i1}^{t}, x_{i2}^{t}\}=\text{Split}(x_{i}^{t})$. The discrete matrices $\bar{A}$ and $\bar{B}$ can be derived from Equation \ref{matrix}.

\textbf{Communication Block}\quad The Communication Block complements the Selection Block by enabling agents to incorporate global information into their decision-making process. In this block, agents prioritize suggestions from other agents and receive information through attention-weighted mechanisms \cite{r36}. Given the hidden states of agents $i$ and $j$ as input, we consider two learnable matrices: the self-query matrix $q_{i}^{t}=W_{q}h_{i}^{t}$ and the cognition matrix $k_{j}^{t}=W_{k}h_{j}^{t}$, where $W_{q}$ , $W_{k}$ are both learnable linear transformations. The calculation of attention weights proceeds as follows:
\begin{equation}
 \begin{aligned}
 c_{i,j}^{t}&=\frac {(q_{i}^{t})^{T} k_{j}^{t}}{\sqrt{d_{h}}}
 \end{aligned}
\end{equation}
\begin{equation}
 \begin{aligned}
 w_{i,j}^{t}&=\frac{\exp(c_{i,j}^{t})}{\sum_{k=1}^{N} \exp(c_{i, k}^{t})}
 \end{aligned}
\end{equation}
$d_{h}$ represents the dimension of the hidden state, and  $\exp$ denotes the exponential operation. Then the true information $v_{i}^{t}$ obtained by agent $i$ can be calculated as $v_{i}^{t} = \sum_{i\neq j} w_{i,j}h_{j}^{t}$.

\begin{figure*}[t]
\begin{center}
\centerline{\includegraphics[scale=0.61]{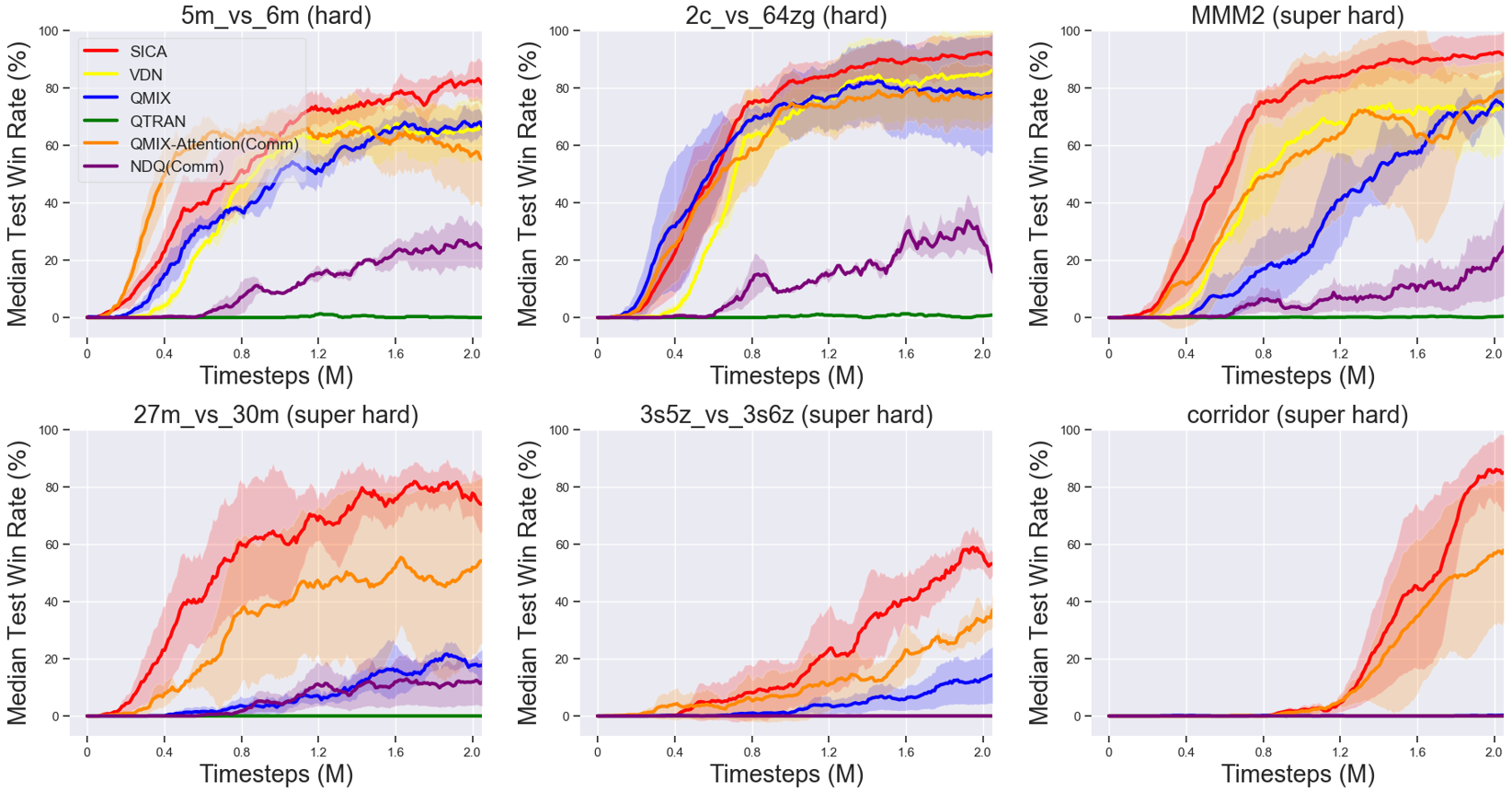}}
\caption{Performance comparison between SICA and baselines on SMAC.}
\label{SMACFIG}
\end{center}
\vskip -0.3in
\end{figure*}

\textbf{Regeneration Block}\quad To ensure that decision-making relies solely on local information, we must convert the existing centralized framework into a decentralized one. Hence, we introduce the Regeneration Block, comprising the Selection Block and an MLP, which exclusively takes the agent $i$'s observation-action pair as input, as depicted in the central section of Figure \ref{Framework}. This Regeneration Block allows us to derive the regenerated information $\hat{v}_{i}$, which continuously approximates the true information $v_{i}$. It's notable that, during the initial stages of training, to ensure accurate learning of global information by the Regeneration Block, we include observation-action pairs from other agents in its minibuffer, gradually reducing it to $0$ over time.

We utilize exponential weighted averaging to ensure that the regenerated information $\hat{v}_{i}^{t}$ converges towards the true information $v_{i}^{t}$. Then, we compute the cross-information $\bar{v}_{i}^{t}$ by weighting both of them.

\begin{equation}
 \begin{aligned}
\bar{v}_{i}^{t}=(1-\alpha(t))\hat{v}^{t}+\alpha(t) v_{i}^{t}
 \end{aligned}
\end{equation}
Here, $\alpha(t)$ is dynamic, and to ensure a smoother transition of the framework, we update it using a method similar to cosine annealing:

\begin{equation}
 \begin{aligned}
\alpha(t)=\alpha_{start}+(\alpha_{final} - \alpha_{start})\cos (\frac{t}{t_{\max}}\pi).
 \end{aligned}
\end{equation}
We set $\alpha_{start}$ to 1 and $\alpha_{final}$ to 0. 

As the training progresses, the cross information $\bar{v}_{i}^{t}$ gradually transitions into regenerated information, allowing the framework to rely less on true information after training completion. Finally, we concatenate the $\bar{v}_{i}^{t}$ with $h_{i}^{t}$ and pass it through an MLP block, obtaining the action value $Q_{i}$ of agent $i$.

Through the Selection Block and Communication Block, agents effectively utilize information, while the Regeneration Block facilitates the transition from centralized to decentralized decision-making. Overall, compared to other CTDE methods, SICA achieves a balance between decision-making capability and generality.

\subsection{Learning Objective}
The overall learning objective of our method is divided into two parts: the TD loss function, which constitutes the end-to-end optimal value decomposition, and the minimization of the regeneration information error. 

The TD loss function part is the same as in QMIX in Equation \ref{TDLOSS} and has the following form:

\begin{equation}
 \begin{aligned}
 \mathcal{L}_{TD}(\theta)=E_{\mathcal{D}}[(r +  \gamma\max\limits_{\boldsymbol{u}^{\prime}} Q^{'}_{tot}(\boldsymbol{\tau}^{\prime},\boldsymbol{u}^{\prime},s^{\prime};\theta^{-})- Q_{tot}(\boldsymbol{\tau},\boldsymbol{u},s;\theta))^{2}]
 \end{aligned}
\end{equation}
Here, $\theta$ represents the learnable parameters, $\theta^{-}$ are the parameters of a target network as in DQN, $E[\cdot]$ denotes the expectation function, and $\mathcal{D}$ represents the replay buffer of transitions.

The part of minimizing the regenerated information error is to ensure that the regenerated information $\hat{v}_{i}$ closely approximates the true information $v_{i}$, facilitating the smooth transition of SICA from a centralized architecture to a decentralized one. To achieve this objective, we introduce an auxiliary loss function called alignment loss function $\mathcal{L}_{Align}$ , formulated as follows:

\begin{equation}
 \begin{aligned}
 \mathcal{L}_{Align}(\theta)=\frac{1}{n}\sum\limits_{i=1}^{n} E[(\hat{v}_{i}(h_{i};\theta^{-})-v_{i}(\boldsymbol{h};\theta))^{2}]
 \end{aligned}
\end{equation}
Where $E[\cdot]$ denotes the expectation function and $n$ denotes the number of agents.

These two parts are trained simultaneously, but with a gradual increase in the importance of the alignment loss function $\mathcal{L}_{Align}$ to ensure the performance of the architecture. The total loss function can be expressed as follows:
\begin{equation}
 \begin{aligned}
 \mathcal{L}_{tot}(\boldsymbol{\tau},\boldsymbol{u},s,h_{i},\boldsymbol{h};\theta)=\mathcal{L}_{TD}+\sigma(t)\mathcal{L}_{Align}
 \end{aligned}
\end{equation}
$t$ denotes a time step, and $\sigma(t)$ is a threshold function defined as follows:
\begin{equation}
\sigma(t)=\left\{
	\begin{aligned}
	\beta_{1} \quad t\leq T\\
	\beta_{2} \quad t > T\\
	\end{aligned}
	\right
	.
\end{equation}
$T$, $\beta_{1}$ and $\beta_{2}$ are hyperparameters, and they are fixed values that need to be satisfied, i.e., the proportion of reconstruction loss increases as the information increases. 

\begin{figure*}[t]
\centering
\includegraphics[scale=0.61]{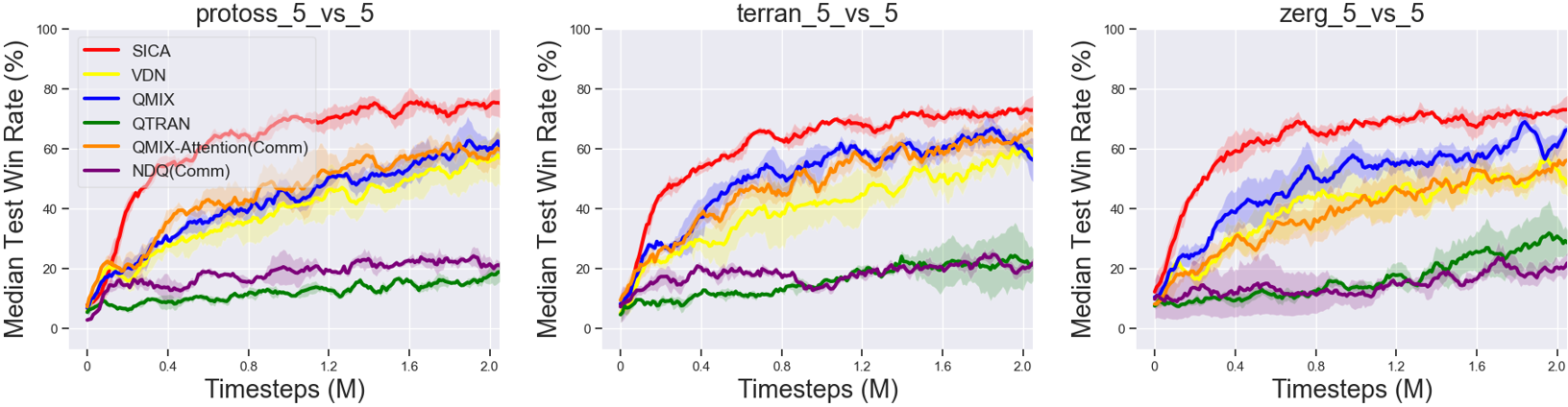}
\caption{Performance comparison between SICA and baselines on SMACv2.}
\label{SMAC255}
\vskip -0.15in
\end{figure*}

\section{Experiments}
In this section, we performed a series of experiments to
determine whether \textbf{1.} SICA outperforms traditional CTDE methods \textbf{2.} SICA outperforms or rivals explicit communication methods. \textbf{3.} SICA outperforms other methods as the volume of information increases. \textbf{4.} Selection mechanism is necessary for SICA to be effective. \textbf{5.} Progressive information regeneration is essential. We conducted these experiments in the StarCraft Multi-Agent Challenge (SMAC)\cite{r31}, SMACv2\cite{r6}, and Google Research Football (GRF)\cite{r9} environments. In all plots, the solid line represents the mean performance over three seeds, and the shaded area denotes the 95$\%$ confidence interval.

\subsection{Performance on Multi-Agent Benchmarks}
In this subsection, our goal is to address questions \textbf{1} and \textbf{2} by assessing SICA's performance across widely-used MARL benchmarks.

\textbf{SMAC}\quad We initiate our evaluation by assessing SICA's performance in the SMAC, where our aim is to control a team of allied units against an enemy team governed by built-in policies. Victory is achieved by eliminating all enemy units within a chapter's time limit. Our metric for evaluation is the alliance team's win rate, which we aim to maximize. We compare SICA against several robust baselines grounded in the CTDE framework.  These include clasSICAl methods like VDN, QMIX, and QTRAN \cite{QTRAN}, as well as communication-based approaches like QMIX-Attention \cite{pymarl2} and NDQ \cite{NDQ}. Given that most methods perform adequately in easy maps, we focus on evaluating their performance in more challenging environments, particularly two hard maps ($5m\_vs\_6m$, $2c\_vs\_64zg$) and four super hard maps ($3s5z\_vs\_3s6z$, $MMM2$, $27m\_\\vs\_30m$, $Corridor$) to offer a comprehensive assessment of SICA's capabilities.

The median win rates across different maps are depicted in Figure \ref{SMACFIG}. SICA consistently outperforms the baselines across all maps, even surpassing explicit communication methods. This underscores the effectiveness of SICA in information processing and highlights the robustness of its information regeneration capability. Across all methods, there is a noticeable decline in win rates as we transition from hard to super hard maps, which aligns with expectations given the heightened complexity of the latter scenarios. It's worth mentioning that QTRAN and NDQ exhibit suboptimal performance. QTRAN exhibits suboptimal performance across all maps, potentially attributable to challenges in credit assignment resulting in the development of passive agents. Meanwhile, NDQ demonstrates efficacy solely on select maps, potentially stemming from instability in its message passing methodology.

\textbf{SMACv2}\quad Next, we evaluate SICA on SMACv2. SMACv2 is an enhanced version of SMAC. In SMACv2, each agent is equipped with randomly generated unit types and initial positions, introducing unpredictability into the environment. These unit types are generated according to a fixed probability distribution, with agents unaware of their own unit types, necessitating adaptable strategies capable of addressing all potential unit types. Figure \ref{SMAC255} presents a comparative analysis between SICA and baseline methods on three randomly generated maps for the $Protoss$, $Terran$, and $Zerg$ races. Remarkably, SICA consistently outperformed the baselines, even in highly stochastic SMACv2 environments. This finding underscores the effectiveness of SICA in task completion, highlighting the robustness of its selection mechanism and information regeneration capabilities.

\textbf{GRF}\quad Finally, we  evaluated SICA's performance on GRF, a MARL benchmark based on the open-source game Gameplay Football. In this environment, agents collaborate to orchestrate attacks, with rewards solely granted upon goal scoring. In Table \ref{GRFTABLE}, we selected the explicit communication method QMIX-Attention and the current state-of-the-art method CDS-QMIX \cite{r20} as baselines. We then compared SICA with these baselines across the two most challenging scenarios. The agents were trained for 10 million steps using 8 threads in all scenarios. The results demonstrate that SICA significantly outperformed the other methods, thereby highlighting its efficacy in diverse environments.

\begin{table}[h]
\vspace{-0.2cm}
  \caption{Performance comparison between SICA and baselines on GRF}
  \vspace{-0.3cm}
  \label{GRFTABLE}
  \centering
  \scalebox{0.85}{
  \begin{tabular}{c|c|c}
\toprule
Algorithm & $3\_vs\_1\_with\_keeper$ & $counterattack\_hard$ \\
\midrule
CDS-QMIX & 0.760 $\pm$ 0.141 & 0.585 $\pm$ 0.072 \\
QMIX-Attention (Comm) & 0.605 $\pm$ 0.182 & 0.397 $\pm$ 0.114 \\
\textbf{SICA (Ours)} & \textbf{0.854} $\pm$ \textbf{0.061} & \textbf{0.671} $\pm$ \textbf{0.141} \\
\bottomrule
  \end{tabular}}
\end{table}
\vspace{-0.3cm}

\begin{figure*}[t]
\begin{center}
\centerline{\includegraphics[scale=0.61]{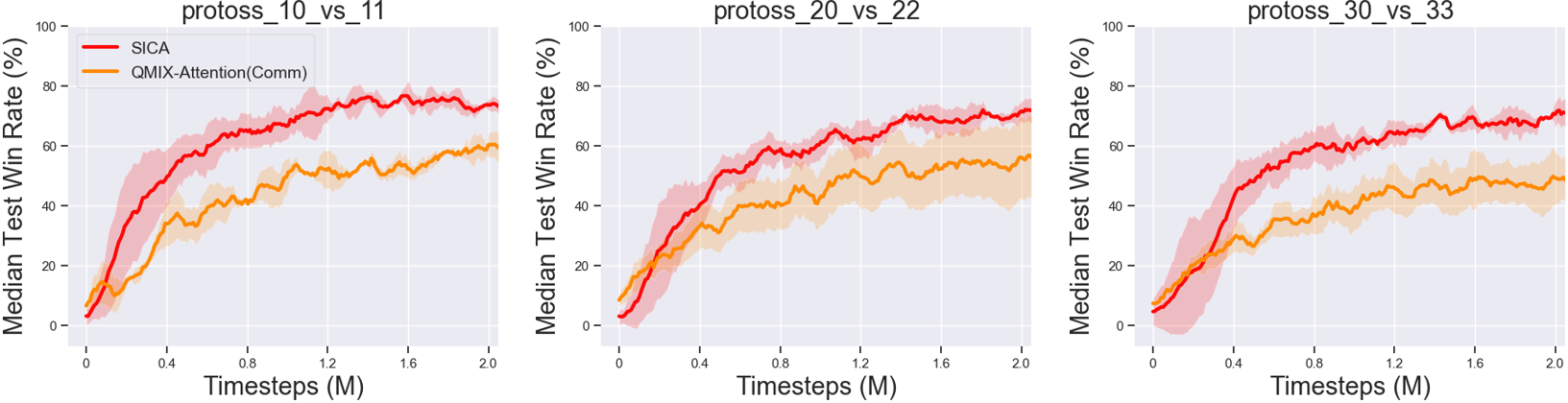}}
\caption{Learning curves with different numbers of agents in SMACv2 $protoss$.}
\label{PPPP}
\end{center}
\end{figure*}

\subsection{Ablation Studies}
In this subsection, we will conduct ablation studies. Three sets of experiments were conducted to address questions \textbf{3}, \textbf{4}, and \textbf{5}, respectively. All experiments were performed in the challenging SMACv2 scenario.

\textbf{Different numbers of agents}\quad Given SICA's ability to encapsulate historical information, we hypothesized that it could effectively handle larger volumes of data, prompting question \textbf{3}. To answer this question, we tested scenarios with different numbers of agents. We compared SICA with the baseline, QMIX-Attention, the best-performing explicit communication method identified in earlier experiments. As illustrated in Figure \ref{PPPP}, SICA consistently outperformed the baseline across all scenarios, highlighting its effectiveness in handling larger-scale multi-agent environments.

\begin{figure}[h!]
\vspace{-0.1cm}
\centering
\includegraphics[scale=0.43]{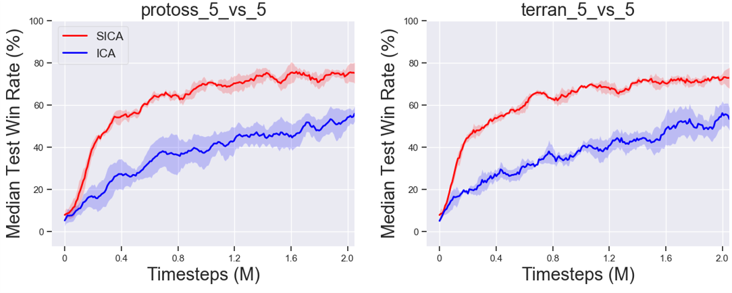}
\caption{Performance comparison
between SICA and ICA.}
\label{ICA}
\vskip -0.12in
\end{figure}

\textbf{Replace Selection Block}\quad To address question \textbf{4}, we introduce a variant of SICA called ICA. In this variant, the Selection is replaced by a gated structure consisting of a MLP Block and a GRU Cell. The results of SICA and ICA in $protoss\_5\_vs\_5$ and $terran\_5\_vs\_5$ scenarios are shown in Figure \ref{ICA}. We observe that the performance of SICA markedly surpassed that of ICA. This suggests that the Selection Block consistently enhances the agents' ability and underscores the significance of information filtering in multi-agent tasks.

\begin{figure}[h!]
\vspace{-0.1cm}
\centering
\includegraphics[scale=0.43]{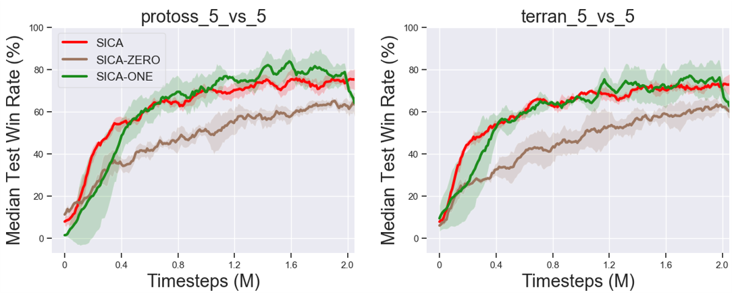}
\caption{Performance comparison
between SICA and SICA's Variants.}
\label{ALPHA}
\vskip -0.12in
\end{figure}

\textbf{Fixed $\alpha$}\quad Does SICA really need progressive information regeneration? To answer question \textbf{5}, we set up two scenarios: one where $\alpha(t)$ is fixed at 0, disregarding $\mathcal{L}_{align}$
 , which we call SICA-ZERO; and another where $\alpha(t)$ is fixed at 1 and suddenly switches to 0 near the end of the training, which we call SICA-ONE. For SICA-ZERO, the communication module does not participate, making SICA equivalent to a traditional CTDE method, relying on local information for training. For SICA-ONE, the training process fully depends on communication, forcing the agents to learn complex information directly within a short time. The comparison between these two scenarios and SICA is shown in Figure \ref{ALPHA}. Regardless of the value of $\alpha(t)$, SICA consistently performs better. Therefore, progressive information regeneration is necessary as it guides the learning process and enhances the agents' capabilities.

\subsection{More Studies}
\begin{figure}[h]
\begin{center}
\centerline{\includegraphics[scale=0.22]{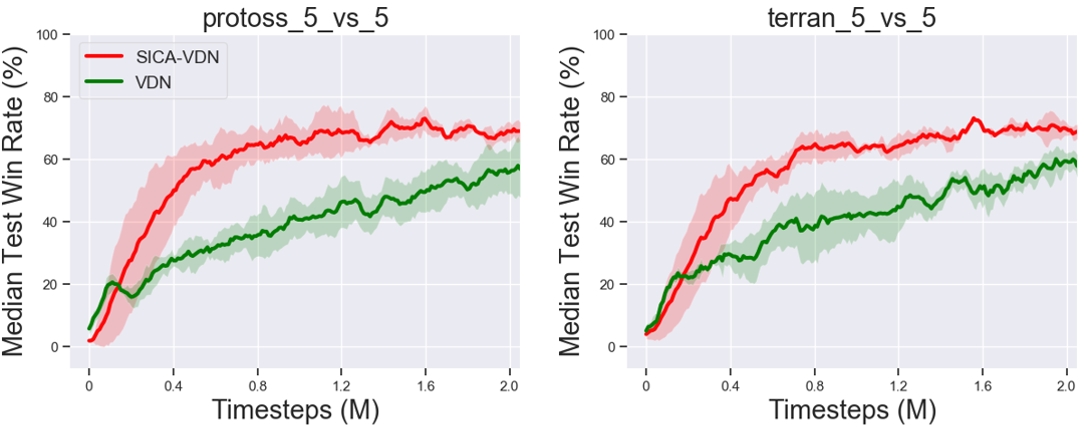}}
\caption{Performance comparison between SICA-VDN and VDN.}
\label{SICAVDN}
\end{center}
\end{figure}

\textbf{Applying SICA to VDN} \quad To demonstrate the versatility of SICA as a plug-and-play framework, we apply SICA to VDN, referred to as SICA-VDN, in this section. The comparison curve between SICA-VDN and VDN is illustrated in Figure \ref{SICAVDN}, showing that SICA markedly improves the performance of the baseline VDN.

\vskip -0.2in

\begin{figure*}[t]
\vskip -0.07in
\begin{center}
\centerline{\includegraphics[scale=0.3]{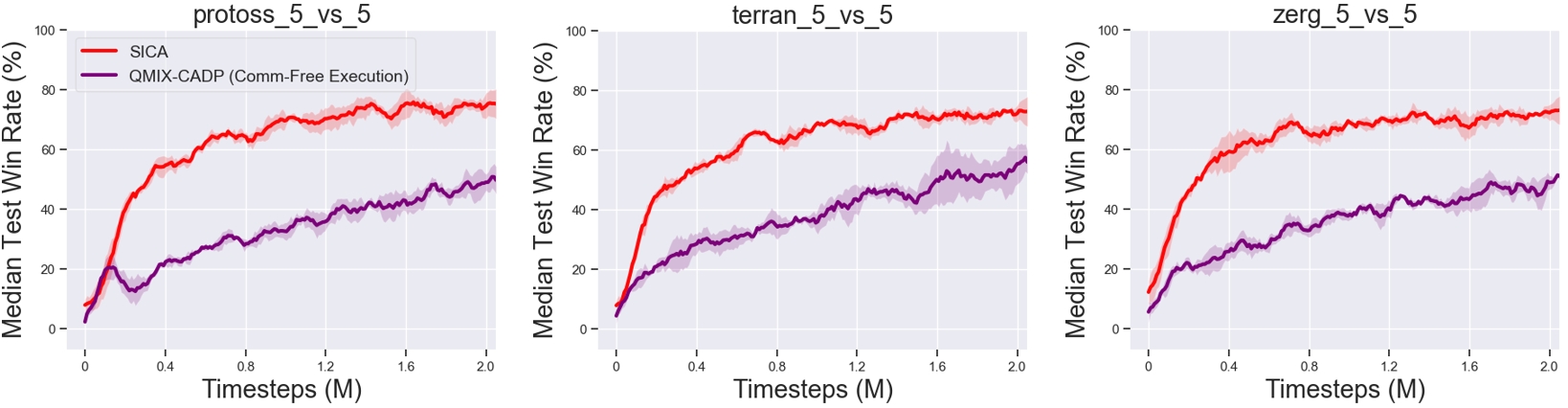}}
\caption{Performance comparison between SICA and QMIX-CADP.}
\label{SICAQMIXCADP}
\end{center}
\vskip -0.2in
\end{figure*}

\textbf{Comparing SICA with another communication-free execution method.} \quad In this section, we compare SICA with another communication-free execution baseline, QMIX-CADP \cite{r44},on SMACv2. QMIX-CADP adopts direct pruning to transition the framework into a decentralized framework. In Figure \ref{SICAQMIXCADP}, we can observe that SICA's performance is significantly better than that of QMIX-CADP, indicating that a gradual framework transition is more reasonable than a direct transition.

\begin{figure*}[t]
	\centering
	\subfloat[Protoss\_3\_vs\_5]{\includegraphics[width=1.1\columnwidth, height=.26\hsize]{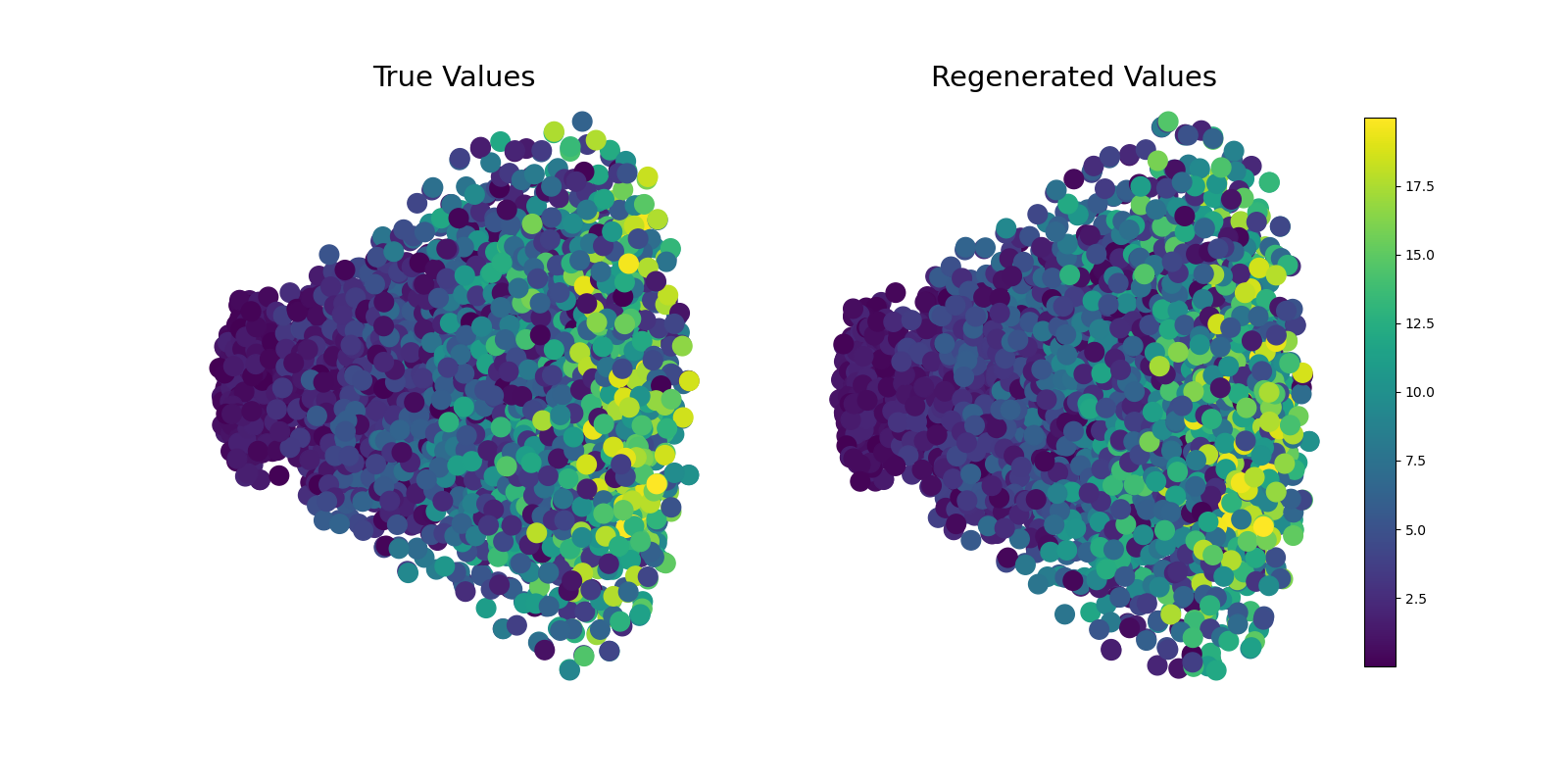}}
	\subfloat[Terran\_3\_vs\_5]{\includegraphics[width=1.1\columnwidth, height=.26\hsize]{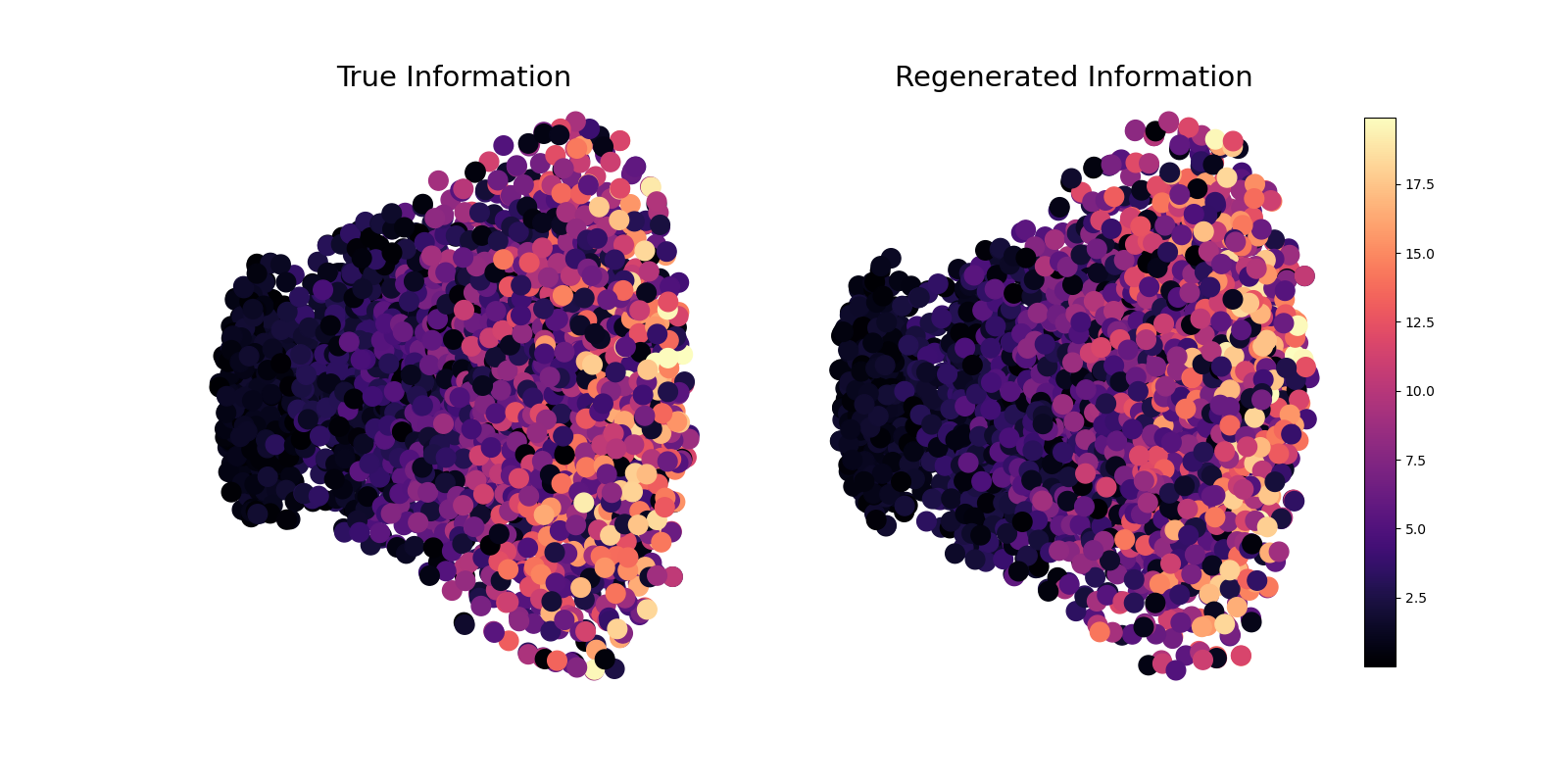}}
	\caption{Visualizations of the regenerated information.}
	\label{visual}
	\vskip -0.15in
\end{figure*}

\textbf{Visualization} \quad To verify whether the Regeneration Block can truly approximate the true information, we visualize the true information and the regenerated information from the $Protoss$ and $Terran$ maps using t-SNE \cite{tSNE} compression into two-dimensional embeddings. In this section, we adopt the $3\_vs\_5$ mode with higher difficulty and randomness to enhance the difficulty of regeneration. As shown in Figure \ref{visual}, although some bias is present in both scenarios, the regenerated information can reflect the distribution of the true information effectively. Therefore, the regeneration process is effective.

\textbf{Scalability} \quad We combined SICA with the actor-critic method MADDPG and compared it to the traditional MADDPG algorithm in the $Predator-Prey$, $Navigation$, and $Pantomime$ environments. Table \ref{3TABLE} presents the experimental results after 2 million training steps, demonstrating that SICA continues to enhance the overall performance of the framework.

\begin{table}[h]
\vspace{-0.2cm}
  \caption{Performance comparison between SICA-MADDPG and MADDPG}

  \label{3TABLE}

  \centering
  \scalebox{0.82}{
  \begin{tabular}{c|c|c|c}
\toprule
Algorithm & $Predator-Prey$ & $Navigation$ & $Pantomime$\\
\midrule
MADDPG & 831.48 $\pm$ 98.73 & -233.31 $\pm$ 32.10 & -812.79 $\pm$ 592.01\\
\textbf{SICA-MADDPG} & \textbf{725.11} $\pm$ \textbf{312.26} & \textbf{-255.23} $\pm$ \textbf{61.42} & \textbf{-926.55} $\pm$ \textbf{601.92}\\
\bottomrule
  \end{tabular}}
\end{table}
\vspace{-0.3cm}

\section{Limitations}
\label{limits}
In this section, we discuss four limitations of SICA. First, the Regeneration Block may struggle when agents' observations or trajectories are dissimilar, although a mini-buffer helps alleviate this issue. Second, our experiments were constrained by computational resources, and further testing in larger environments is needed. Third, applying SICA to new tasks requires tuning several hyperparameters. Finally, parameter sharing is used to accelerate training; without it, training times would be significantly longer.

\section{Conclusion}
In this paper, we introduced a novel MARL architecture named SICA, designed to enhance agents' information handling capabilities and improve the framework's generality. By integrating information selection with communication mechanisms, SICA empowers agents to autonomously choose relevant information while incorporating information from other agents. To accommodate to communication-limited environments, SICA gradually learns the tacit understanding between agents, eventually transitioning to a fully decentralized framework. Experimental results illustrate SICA's effectiveness in regenerating global information and significantly enhancing performance in challenging multi-agent tasks through information selection. In future work, we aim to seamlessly extend the framework to encompass various CTDE methods.

\balance





\newpage
\bibliographystyle{ACM-Reference-Format} 
\bibliography{example_paper}


\begin{thebibliography}{45}


\ifx \showCODEN    \undefined \def \showCODEN     #1{\unskip}     \fi
\ifx \showDOI      \undefined \def \showDOI       #1{#1}\fi
\ifx \showISBNx    \undefined \def \showISBNx     #1{\unskip}     \fi
\ifx \showISBNxiii \undefined \def \showISBNxiii  #1{\unskip}     \fi
\ifx \showISSN     \undefined \def \showISSN      #1{\unskip}     \fi
\ifx \showLCCN     \undefined \def \showLCCN      #1{\unskip}     \fi
\ifx \shownote     \undefined \def \shownote      #1{#1}          \fi
\ifx \showarticletitle \undefined \def \showarticletitle #1{#1}   \fi
\ifx \showURL      \undefined \def \showURL       {\relax}        \fi
\providecommand\bibfield[2]{#2}
\providecommand\bibinfo[2]{#2}
\providecommand\natexlab[1]{#1}
\providecommand\showeprint[2][]{arXiv:#2}

\bibitem[\protect\citeauthoryear{Berner, Brockman, Chan, Cheung, Debiak,
  Dennison, Farhi, Fischer, Hashme, Hesse, J{\'o}zefowicz, Gray, Olsson,
  Pachocki, Petrov, de~Oliveira~Pinto, Raiman, Salimans, Schlatter, Schneider,
  Sidor, Sutskever, Tang, Wolski, and Zhang}{Berner et~al\mbox{.}}{2019}]%
        {r1}
\bibfield{author}{\bibinfo{person}{Christopher Berner}, \bibinfo{person}{Greg
  Brockman}, \bibinfo{person}{Brooke Chan}, \bibinfo{person}{Vicki Cheung},
  \bibinfo{person}{Przemyslaw Debiak}, \bibinfo{person}{Christy Dennison},
  \bibinfo{person}{David Farhi}, \bibinfo{person}{Quirin Fischer},
  \bibinfo{person}{Shariq Hashme}, \bibinfo{person}{Christopher Hesse},
  \bibinfo{person}{Rafal J{\'o}zefowicz}, \bibinfo{person}{Scott Gray},
  \bibinfo{person}{Catherine Olsson}, \bibinfo{person}{Jakub~W. Pachocki},
  \bibinfo{person}{Michael Petrov}, \bibinfo{person}{Henrique~Pond{\'e} de
  Oliveira~Pinto}, \bibinfo{person}{Jonathan Raiman}, \bibinfo{person}{Tim
  Salimans}, \bibinfo{person}{Jeremy Schlatter}, \bibinfo{person}{Jonas
  Schneider}, \bibinfo{person}{Szymon Sidor}, \bibinfo{person}{Ilya Sutskever},
  \bibinfo{person}{Jie Tang}, \bibinfo{person}{Filip Wolski}, {and}
  \bibinfo{person}{Susan Zhang}.} \bibinfo{year}{2019}\natexlab{}.
\newblock \showarticletitle{Dota 2 with Large Scale Deep Reinforcement
  Learning}.
\newblock \bibinfo{journal}{\emph{ArXiv}}  \bibinfo{volume}{abs/1912.06680}
  (\bibinfo{year}{2019}).
\newblock


\bibitem[\protect\citeauthoryear{Bernstein, Zilberstein, and
  Immerman}{Bernstein et~al\mbox{.}}{2000}]%
        {r2}
\bibfield{author}{\bibinfo{person}{Daniel~S. Bernstein},
  \bibinfo{person}{Shlomo Zilberstein}, {and} \bibinfo{person}{Neil Immerman}.}
  \bibinfo{year}{2000}\natexlab{}.
\newblock \showarticletitle{The Complexity of Decentralized Control of Markov
  Decision Processes}.
\newblock \bibinfo{journal}{\emph{ArXiv}}  \bibinfo{volume}{abs/1301.3836}
  (\bibinfo{year}{2000}).
\newblock


\bibitem[\protect\citeauthoryear{Claus and Boutilier}{Claus and
  Boutilier}{1998}]%
        {r3}
\bibfield{author}{\bibinfo{person}{Caroline Claus} {and} \bibinfo{person}{Craig
  Boutilier}.} \bibinfo{year}{1998}\natexlab{}.
\newblock \showarticletitle{The Dynamics of Reinforcement Learning in
  Cooperative Multiagent Systems}. In \bibinfo{booktitle}{\emph{AAAI/IAAI}}.
\newblock


\bibitem[\protect\citeauthoryear{Das, Gervet, Romoff, Batra, Parikh, Rabbat,
  and Pineau}{Das et~al\mbox{.}}{2018}]%
        {r4}
\bibfield{author}{\bibinfo{person}{Abhishek Das},
  \bibinfo{person}{Th{\'e}ophile Gervet}, \bibinfo{person}{Joshua Romoff},
  \bibinfo{person}{Dhruv Batra}, \bibinfo{person}{Devi Parikh},
  \bibinfo{person}{Michael~G. Rabbat}, {and} \bibinfo{person}{Joelle Pineau}.}
  \bibinfo{year}{2018}\natexlab{}.
\newblock \showarticletitle{TarMAC: Targeted Multi-Agent Communication}. In
  \bibinfo{booktitle}{\emph{International Conference on Machine Learning}}.
\newblock


\bibitem[\protect\citeauthoryear{Ding, Huang, and Lu}{Ding
  et~al\mbox{.}}{2020}]%
        {r5}
\bibfield{author}{\bibinfo{person}{Ziluo Ding}, \bibinfo{person}{Tiejun Huang},
  {and} \bibinfo{person}{Zongqing Lu}.} \bibinfo{year}{2020}\natexlab{}.
\newblock \showarticletitle{Learning Individually Inferred Communication for
  Multi-Agent Cooperation}.
\newblock \bibinfo{journal}{\emph{ArXiv}}  \bibinfo{volume}{abs/2006.06455}
  (\bibinfo{year}{2020}).
\newblock


\bibitem[\protect\citeauthoryear{Ellis, Moalla, Samvelyan, Sun, Mahajan,
  Foerster, and Whiteson}{Ellis et~al\mbox{.}}{2022}]%
        {r6}
\bibfield{author}{\bibinfo{person}{Benjamin Ellis}, \bibinfo{person}{S.
  Moalla}, \bibinfo{person}{Mikayel Samvelyan}, \bibinfo{person}{Mingfei Sun},
  \bibinfo{person}{Anuj Mahajan}, \bibinfo{person}{Jakob~Nicolaus Foerster},
  {and} \bibinfo{person}{Shimon Whiteson}.} \bibinfo{year}{2022}\natexlab{}.
\newblock \showarticletitle{SMACv2: An Improved Benchmark for Cooperative
  Multi-Agent Reinforcement Learning}.
\newblock \bibinfo{journal}{\emph{ArXiv}}  \bibinfo{volume}{abs/2212.07489}
  (\bibinfo{year}{2022}).
\newblock


\bibitem[\protect\citeauthoryear{Foerster, Assael, de~Freitas, and
  Whiteson}{Foerster et~al\mbox{.}}{2016}]%
        {r7}
\bibfield{author}{\bibinfo{person}{Jakob~N. Foerster}, \bibinfo{person}{Yannis
  Assael}, \bibinfo{person}{Nando de Freitas}, {and} \bibinfo{person}{Shimon
  Whiteson}.} \bibinfo{year}{2016}\natexlab{}.
\newblock \showarticletitle{Learning to Communicate with Deep Multi-Agent
  Reinforcement Learning}.
\newblock \bibinfo{journal}{\emph{ArXiv}}  \bibinfo{volume}{abs/1605.06676}
  (\bibinfo{year}{2016}).
\newblock


\bibitem[\protect\citeauthoryear{Foerster, Farquhar, Afouras, Nardelli, and
  Whiteson}{Foerster et~al\mbox{.}}{2017}]%
        {r8}
\bibfield{author}{\bibinfo{person}{Jakob~N. Foerster}, \bibinfo{person}{Gregory
  Farquhar}, \bibinfo{person}{Triantafyllos Afouras}, \bibinfo{person}{Nantas
  Nardelli}, {and} \bibinfo{person}{Shimon Whiteson}.}
  \bibinfo{year}{2017}\natexlab{}.
\newblock \showarticletitle{Counterfactual Multi-Agent Policy Gradients}. In
  \bibinfo{booktitle}{\emph{AAAI Conference on Artificial Intelligence}}.
\newblock


\bibitem[\protect\citeauthoryear{Goel, Gu, Donahue, and R'e}{Goel
  et~al\mbox{.}}{2022}]%
        {r9}
\bibfield{author}{\bibinfo{person}{Karan Goel}, \bibinfo{person}{Albert Gu},
  \bibinfo{person}{Chris Donahue}, {and} \bibinfo{person}{Christopher R'e}.}
  \bibinfo{year}{2022}\natexlab{}.
\newblock \showarticletitle{It's Raw! Audio Generation with State-Space
  Models}. In \bibinfo{booktitle}{\emph{International Conference on Machine
  Learning}}.
\newblock


\bibitem[\protect\citeauthoryear{Gu and Dao}{Gu and Dao}{2023}]%
        {r13}
\bibfield{author}{\bibinfo{person}{Albert Gu} {and} \bibinfo{person}{Tri Dao}.}
  \bibinfo{year}{2023}\natexlab{}.
\newblock \showarticletitle{Mamba: Linear-Time Sequence Modeling with Selective
  State Spaces}.
\newblock \bibinfo{journal}{\emph{ArXiv}}  \bibinfo{volume}{abs/2312.00752}
  (\bibinfo{year}{2023}).
\newblock


\bibitem[\protect\citeauthoryear{Gu, Dao, Ermon, Rudra, and R{\'e}}{Gu
  et~al\mbox{.}}{2020}]%
        {r10}
\bibfield{author}{\bibinfo{person}{Albert Gu}, \bibinfo{person}{Tri Dao},
  \bibinfo{person}{Stefano Ermon}, \bibinfo{person}{Atri Rudra}, {and}
  \bibinfo{person}{Christopher R{\'e}}.} \bibinfo{year}{2020}\natexlab{}.
\newblock \showarticletitle{HiPPO: Recurrent Memory with Optimal Polynomial
  Projections}.
\newblock \bibinfo{journal}{\emph{ArXiv}}  \bibinfo{volume}{abs/2008.07669}
  (\bibinfo{year}{2020}).
\newblock


\bibitem[\protect\citeauthoryear{Gu, Goel, and R'e}{Gu et~al\mbox{.}}{2021}]%
        {r11}
\bibfield{author}{\bibinfo{person}{Albert Gu}, \bibinfo{person}{Karan Goel},
  {and} \bibinfo{person}{Christopher R'e}.} \bibinfo{year}{2021}\natexlab{}.
\newblock \showarticletitle{Efficiently Modeling Long Sequences with Structured
  State Spaces}.
\newblock \bibinfo{journal}{\emph{ArXiv}}  \bibinfo{volume}{abs/2111.00396}
  (\bibinfo{year}{2021}).
\newblock


\bibitem[\protect\citeauthoryear{Gupta and Berant}{Gupta and Berant}{2022}]%
        {r14}
\bibfield{author}{\bibinfo{person}{Ankit Gupta} {and} \bibinfo{person}{Jonathan
  Berant}.} \bibinfo{year}{2022}\natexlab{}.
\newblock \showarticletitle{Diagonal State Spaces are as Effective as
  Structured State Spaces}.
\newblock \bibinfo{journal}{\emph{ArXiv}}  \bibinfo{volume}{abs/2203.14343}
  (\bibinfo{year}{2022}).
\newblock


\bibitem[\protect\citeauthoryear{Hu, Jiang, Harding, Wu, and Liao}{Hu
  et~al\mbox{.}}{2021}]%
        {pymarl2}
\bibfield{author}{\bibinfo{person}{Jian Hu}, \bibinfo{person}{Siyang Jiang},
  \bibinfo{person}{Seth~Austin Harding}, \bibinfo{person}{Haibin Wu}, {and}
  \bibinfo{person}{Shihua Liao}.} \bibinfo{year}{2021}\natexlab{}.
\newblock \showarticletitle{Rethinking the Implementation Tricks and
  Monotonicity Constraint in Cooperative Multi-Agent Reinforcement Learning}.
\newblock


\bibitem[\protect\citeauthoryear{Jaques, Lazaridou, Hughes, Çaglar
  G{\"u}lçehre, Ortega, Strouse, Leibo, and de~Freitas}{Jaques
  et~al\mbox{.}}{2018}]%
        {r16}
\bibfield{author}{\bibinfo{person}{Natasha Jaques}, \bibinfo{person}{Angeliki
  Lazaridou}, \bibinfo{person}{Edward Hughes}, \bibinfo{person}{Çaglar
  G{\"u}lçehre}, \bibinfo{person}{Pedro~A. Ortega}, \bibinfo{person}{DJ
  Strouse}, \bibinfo{person}{Joel~Z. Leibo}, {and} \bibinfo{person}{Nando de
  Freitas}.} \bibinfo{year}{2018}\natexlab{}.
\newblock \showarticletitle{Social Influence as Intrinsic Motivation for
  Multi-Agent Deep Reinforcement Learning}. In
  \bibinfo{booktitle}{\emph{International Conference on Machine Learning}}.
\newblock


\bibitem[\protect\citeauthoryear{Jiang and Lu}{Jiang and Lu}{2018}]%
        {r17}
\bibfield{author}{\bibinfo{person}{Jiechuan Jiang} {and}
  \bibinfo{person}{Zongqing Lu}.} \bibinfo{year}{2018}\natexlab{}.
\newblock \showarticletitle{Learning Attentional Communication for Multi-Agent
  Cooperation}. In \bibinfo{booktitle}{\emph{Neural Information Processing
  Systems}}.
\newblock


\bibitem[\protect\citeauthoryear{Knepper, Mavrogiannis, Proft, and
  Liang}{Knepper et~al\mbox{.}}{2017}]%
        {r18}
\bibfield{author}{\bibinfo{person}{Ross~A. Knepper},
  \bibinfo{person}{Christoforos Mavrogiannis}, \bibinfo{person}{Julia Proft},
  {and} \bibinfo{person}{Claire Liang}.} \bibinfo{year}{2017}\natexlab{}.
\newblock \showarticletitle{Implicit Communication in a Joint Action}.
\newblock \bibinfo{journal}{\emph{2017 12th ACM/IEEE International Conference
  on Human-Robot Interaction (HRI}} (\bibinfo{year}{2017}),
  \bibinfo{pages}{283--292}.
\newblock


\bibitem[\protect\citeauthoryear{Kurach, Raichuk, Stańczyk, Zajac, Bachem,
  Espeholt, Riquelme, Vincent, Michalski, Bousquet, and Gelly}{Kurach
  et~al\mbox{.}}{2019}]%
        {r19}
\bibfield{author}{\bibinfo{person}{Karol Kurach}, \bibinfo{person}{Anton
  Raichuk}, \bibinfo{person}{Piotr Stańczyk}, \bibinfo{person}{Michal Zajac},
  \bibinfo{person}{Olivier Bachem}, \bibinfo{person}{Lasse Espeholt},
  \bibinfo{person}{Carlos Riquelme}, \bibinfo{person}{Damien Vincent},
  \bibinfo{person}{Marcin Michalski}, \bibinfo{person}{Olivier Bousquet}, {and}
  \bibinfo{person}{Sylvain Gelly}.} \bibinfo{year}{2019}\natexlab{}.
\newblock \showarticletitle{Google Research Football: A Novel Reinforcement
  Learning Environment}.
\newblock \bibinfo{journal}{\emph{ArXiv}}  \bibinfo{volume}{abs/1907.11180}
  (\bibinfo{year}{2019}).
\newblock


\bibitem[\protect\citeauthoryear{Li, Wu, Wang, Yang, Zhao, and Zhang}{Li
  et~al\mbox{.}}{2021}]%
        {r20}
\bibfield{author}{\bibinfo{person}{Chenghao Li}, \bibinfo{person}{Chengjie Wu},
  \bibinfo{person}{Tonghan Wang}, \bibinfo{person}{Jun Yang},
  \bibinfo{person}{Qianchuan Zhao}, {and} \bibinfo{person}{Chongjie Zhang}.}
  \bibinfo{year}{2021}\natexlab{}.
\newblock \showarticletitle{Celebrating Diversity in Shared Multi-Agent
  Reinforcement Learning}. In \bibinfo{booktitle}{\emph{Neural Information
  Processing Systems}}.
\newblock


\bibitem[\protect\citeauthoryear{Li, Xu, Zhang, and Fan}{Li
  et~al\mbox{.}}{2023}]%
        {r21}
\bibfield{author}{\bibinfo{person}{Dapeng Li}, \bibinfo{person}{Zhiwei Xu},
  \bibinfo{person}{Bin Zhang}, {and} \bibinfo{person}{Guoliang Fan}.}
  \bibinfo{year}{2023}\natexlab{}.
\newblock \showarticletitle{From Explicit Communication to Tacit Cooperation: A
  Novel Paradigm for Cooperative MARL}.
\newblock \bibinfo{journal}{\emph{ArXiv}}  \bibinfo{volume}{abs/2304.14656}
  (\bibinfo{year}{2023}).
\newblock


\bibitem[\protect\citeauthoryear{Liu, Dai, So, and Le}{Liu
  et~al\mbox{.}}{2021}]%
        {gMLP}
\bibfield{author}{\bibinfo{person}{Hanxiao Liu}, \bibinfo{person}{Zihang Dai},
  \bibinfo{person}{David So}, {and} \bibinfo{person}{Quoc~V Le}.}
  \bibinfo{year}{2021}\natexlab{}.
\newblock \showarticletitle{Pay attention to mlps}.
\newblock \bibinfo{journal}{\emph{Advances in neural information processing
  systems}}  \bibinfo{volume}{34} (\bibinfo{year}{2021}),
  \bibinfo{pages}{9204--9215}.
\newblock


\bibitem[\protect\citeauthoryear{Lowe, Wu, Tamar, Harb, Abbeel, and
  Mordatch}{Lowe et~al\mbox{.}}{2017}]%
        {r23}
\bibfield{author}{\bibinfo{person}{Ryan Lowe}, \bibinfo{person}{Yi Wu},
  \bibinfo{person}{Aviv Tamar}, \bibinfo{person}{Jean Harb},
  \bibinfo{person}{P. Abbeel}, {and} \bibinfo{person}{Igor Mordatch}.}
  \bibinfo{year}{2017}\natexlab{}.
\newblock \showarticletitle{Multi-Agent Actor-Critic for Mixed
  Cooperative-Competitive Environments}.
\newblock \bibinfo{journal}{\emph{ArXiv}}  \bibinfo{volume}{abs/1706.02275}
  (\bibinfo{year}{2017}).
\newblock


\bibitem[\protect\citeauthoryear{Mguni, Jennings, and de~Cote}{Mguni
  et~al\mbox{.}}{2018}]%
        {r25}
\bibfield{author}{\bibinfo{person}{David~Henry Mguni}, \bibinfo{person}{Joel
  Jennings}, {and} \bibinfo{person}{Enrique~Munoz de Cote}.}
  \bibinfo{year}{2018}\natexlab{}.
\newblock \showarticletitle{Decentralised Learning in Systems with Many, Many
  Strategic Agents}.
\newblock \bibinfo{journal}{\emph{ArXiv}}  \bibinfo{volume}{abs/1803.05028}
  (\bibinfo{year}{2018}).
\newblock


\bibitem[\protect\citeauthoryear{Mguni, Jennings, Macua, Sison, Ceppi, and
  de~Cote}{Mguni et~al\mbox{.}}{2019}]%
        {r26}
\bibfield{author}{\bibinfo{person}{David~Henry Mguni}, \bibinfo{person}{Joel
  Jennings}, \bibinfo{person}{Sergio~Valcarcel Macua}, \bibinfo{person}{Emilio
  Sison}, \bibinfo{person}{Sofia Ceppi}, {and} \bibinfo{person}{Enrique~Munoz
  de Cote}.} \bibinfo{year}{2019}\natexlab{}.
\newblock \showarticletitle{Coordinating the Crowd: Inducing Desirable
  Equilibria in Non-Cooperative Systems}. In \bibinfo{booktitle}{\emph{Adaptive
  Agents and Multi-Agent Systems}}.
\newblock


\bibitem[\protect\citeauthoryear{Peng, Wen, Yang, Yuan, Tang, Long, and
  Wang}{Peng et~al\mbox{.}}{2017}]%
        {r27}
\bibfield{author}{\bibinfo{person}{Peng Peng}, \bibinfo{person}{Ying Wen},
  \bibinfo{person}{Yaodong Yang}, \bibinfo{person}{Quan Yuan},
  \bibinfo{person}{Zhenkun Tang}, \bibinfo{person}{Haitao Long}, {and}
  \bibinfo{person}{Jun Wang}.} \bibinfo{year}{2017}\natexlab{}.
\newblock \showarticletitle{Multiagent Bidirectionally-Coordinated Nets:
  Emergence of Human-level Coordination in Learning to Play StarCraft Combat
  Games}.
\newblock \bibinfo{journal}{\emph{arXiv: Artificial Intelligence}}
  (\bibinfo{year}{2017}).
\newblock


\bibitem[\protect\citeauthoryear{Qiu, Wang, Dong, Wang, and Strbac}{Qiu
  et~al\mbox{.}}{2023}]%
        {r29}
\bibfield{author}{\bibinfo{person}{Dawei Qiu}, \bibinfo{person}{Jianhong Wang},
  \bibinfo{person}{Zihang Dong}, \bibinfo{person}{Yi Wang}, {and}
  \bibinfo{person}{Goran Strbac}.} \bibinfo{year}{2023}\natexlab{}.
\newblock \showarticletitle{Mean-Field Multi-Agent Reinforcement Learning for
  Peer-to-Peer Multi-Energy Trading}.
\newblock \bibinfo{journal}{\emph{IEEE Transactions on Power Systems}}
  \bibinfo{volume}{38} (\bibinfo{year}{2023}), \bibinfo{pages}{4853--4866}.
\newblock


\bibitem[\protect\citeauthoryear{Qiu, Wang, Wang, and Strbac}{Qiu
  et~al\mbox{.}}{2021}]%
        {r28}
\bibfield{author}{\bibinfo{person}{Dawei Qiu}, \bibinfo{person}{Jianhong Wang},
  \bibinfo{person}{Junkai Wang}, {and} \bibinfo{person}{Goran Strbac}.}
  \bibinfo{year}{2021}\natexlab{}.
\newblock \showarticletitle{Multi-Agent Reinforcement Learning for Automated
  Peer-to-Peer Energy Trading in Double-Side Auction Market}. In
  \bibinfo{booktitle}{\emph{International Joint Conference on Artificial
  Intelligence}}.
\newblock


\bibitem[\protect\citeauthoryear{Rashid, Samvelyan, Witt, Farquhar, Foerster,
  and Whiteson}{Rashid et~al\mbox{.}}{2018}]%
        {r30}
\bibfield{author}{\bibinfo{person}{Tabish Rashid}, \bibinfo{person}{Mikayel
  Samvelyan}, \bibinfo{person}{C.~S.~D. Witt}, \bibinfo{person}{Gregory
  Farquhar}, \bibinfo{person}{Jakob~N. Foerster}, {and} \bibinfo{person}{Shimon
  Whiteson}.} \bibinfo{year}{2018}\natexlab{}.
\newblock \showarticletitle{QMIX: Monotonic Value Function Factorisation for
  Deep Multi-Agent Reinforcement Learning}.
\newblock \bibinfo{journal}{\emph{ArXiv}}  \bibinfo{volume}{abs/1803.11485}
  (\bibinfo{year}{2018}).
\newblock


\bibitem[\protect\citeauthoryear{Samvelyan, Rashid, Witt, Farquhar, Nardelli,
  Rudner, Hung, Torr, Foerster, and Whiteson}{Samvelyan et~al\mbox{.}}{2019}]%
        {r31}
\bibfield{author}{\bibinfo{person}{Mikayel Samvelyan}, \bibinfo{person}{Tabish
  Rashid}, \bibinfo{person}{C.~S.~D. Witt}, \bibinfo{person}{Gregory Farquhar},
  \bibinfo{person}{Nantas Nardelli}, \bibinfo{person}{Tim G.~J. Rudner},
  \bibinfo{person}{Chia-Man Hung}, \bibinfo{person}{Philip H.~S. Torr},
  \bibinfo{person}{Jakob~N. Foerster}, {and} \bibinfo{person}{Shimon
  Whiteson}.} \bibinfo{year}{2019}\natexlab{}.
\newblock \showarticletitle{The StarCraft Multi-Agent Challenge}.
\newblock \bibinfo{journal}{\emph{ArXiv}}  \bibinfo{volume}{abs/1902.04043}
  (\bibinfo{year}{2019}).
\newblock


\bibitem[\protect\citeauthoryear{Son, Kim, Kang, Hostallero, and Yi}{Son
  et~al\mbox{.}}{2019}]%
        {QTRAN}
\bibfield{author}{\bibinfo{person}{Kyunghwan Son}, \bibinfo{person}{Daewoo
  Kim}, \bibinfo{person}{Wan~Ju Kang}, \bibinfo{person}{David~Earl Hostallero},
  {and} \bibinfo{person}{Yung Yi}.} \bibinfo{year}{2019}\natexlab{}.
\newblock \showarticletitle{QTRAN: Learning to Factorize with Transformation
  for Cooperative Multi-Agent Reinforcement Learning}.
\newblock \bibinfo{journal}{\emph{ArXiv}}  \bibinfo{volume}{abs/1905.05408}
  (\bibinfo{year}{2019}).
\newblock


\bibitem[\protect\citeauthoryear{Sukhbaatar, Szlam, and Fergus}{Sukhbaatar
  et~al\mbox{.}}{2016}]%
        {r32}
\bibfield{author}{\bibinfo{person}{Sainbayar Sukhbaatar},
  \bibinfo{person}{Arthur Szlam}, {and} \bibinfo{person}{Rob Fergus}.}
  \bibinfo{year}{2016}\natexlab{}.
\newblock \showarticletitle{Learning Multiagent Communication with
  Backpropagation}. In \bibinfo{booktitle}{\emph{Neural Information Processing
  Systems}}.
\newblock


\bibitem[\protect\citeauthoryear{Sunehag, Lever, Gruslys, Czarnecki, Zambaldi,
  Jaderberg, Lanctot, Sonnerat, Leibo, Tuyls, and Graepel}{Sunehag
  et~al\mbox{.}}{2017}]%
        {r33}
\bibfield{author}{\bibinfo{person}{Peter Sunehag}, \bibinfo{person}{Guy Lever},
  \bibinfo{person}{Audrunas Gruslys}, \bibinfo{person}{Wojciech~M. Czarnecki},
  \bibinfo{person}{Vin{\'i}cius~Flores Zambaldi}, \bibinfo{person}{Max
  Jaderberg}, \bibinfo{person}{Marc Lanctot}, \bibinfo{person}{Nicolas
  Sonnerat}, \bibinfo{person}{Joel~Z. Leibo}, \bibinfo{person}{Karl Tuyls},
  {and} \bibinfo{person}{Thore Graepel}.} \bibinfo{year}{2017}\natexlab{}.
\newblock \showarticletitle{Value-Decomposition Networks For Cooperative
  Multi-Agent Learning}.
\newblock \bibinfo{journal}{\emph{ArXiv}}  \bibinfo{volume}{abs/1706.05296}
  (\bibinfo{year}{2017}).
\newblock


\bibitem[\protect\citeauthoryear{Tan}{Tan}{1993}]%
        {dec}
\bibfield{author}{\bibinfo{person}{Ming Tan}.} \bibinfo{year}{1993}\natexlab{}.
\newblock \showarticletitle{Multi-agent reinforcement learning: Independent vs.
  cooperative agents}. In \bibinfo{booktitle}{\emph{Proceedings of the tenth
  international conference on machine learning}}. \bibinfo{pages}{330--337}.
\newblock


\bibitem[\protect\citeauthoryear{Tian, Zou, Davies, Warr, Wu, Bou-Ammar, and
  Wang}{Tian et~al\mbox{.}}{2018}]%
        {r35}
\bibfield{author}{\bibinfo{person}{Zheng Tian}, \bibinfo{person}{Shihao Zou},
  \bibinfo{person}{Ian Davies}, \bibinfo{person}{Tim Warr},
  \bibinfo{person}{Lisheng Wu}, \bibinfo{person}{Haitham Bou-Ammar}, {and}
  \bibinfo{person}{Jun Wang}.} \bibinfo{year}{2018}\natexlab{}.
\newblock \showarticletitle{Learning to Communicate Implicitly by Actions}. In
  \bibinfo{booktitle}{\emph{AAAI Conference on Artificial Intelligence}}.
\newblock


\bibitem[\protect\citeauthoryear{Van~der Maaten and Hinton}{Van~der Maaten and
  Hinton}{2008}]%
        {tSNE}
\bibfield{author}{\bibinfo{person}{Laurens Van~der Maaten} {and}
  \bibinfo{person}{Geoffrey Hinton}.} \bibinfo{year}{2008}\natexlab{}.
\newblock \showarticletitle{Visualizing data using t-SNE.}
\newblock \bibinfo{journal}{\emph{Journal of machine learning research}}
  \bibinfo{volume}{9}, \bibinfo{number}{11} (\bibinfo{year}{2008}).
\newblock


\bibitem[\protect\citeauthoryear{Vaswani, Shazeer, Parmar, Uszkoreit, Jones,
  Gomez, Kaiser, and Polosukhin}{Vaswani et~al\mbox{.}}{2017}]%
        {r36}
\bibfield{author}{\bibinfo{person}{Ashish Vaswani}, \bibinfo{person}{Noam~M.
  Shazeer}, \bibinfo{person}{Niki Parmar}, \bibinfo{person}{Jakob Uszkoreit},
  \bibinfo{person}{Llion Jones}, \bibinfo{person}{Aidan~N. Gomez},
  \bibinfo{person}{Lukasz Kaiser}, {and} \bibinfo{person}{Illia Polosukhin}.}
  \bibinfo{year}{2017}\natexlab{}.
\newblock \showarticletitle{Attention is All you Need}. In
  \bibinfo{booktitle}{\emph{Neural Information Processing Systems}}.
\newblock


\bibitem[\protect\citeauthoryear{Wang, Ren, Liu, Yu, and Zhang}{Wang
  et~al\mbox{.}}{2020}]%
        {r37}
\bibfield{author}{\bibinfo{person}{Jianhao Wang}, \bibinfo{person}{Zhizhou
  Ren}, \bibinfo{person}{Terry Liu}, \bibinfo{person}{Yang Yu}, {and}
  \bibinfo{person}{Chongjie Zhang}.} \bibinfo{year}{2020}\natexlab{}.
\newblock \showarticletitle{QPLEX: Duplex Dueling Multi-Agent Q-Learning}.
\newblock \bibinfo{journal}{\emph{ArXiv}}  \bibinfo{volume}{abs/2008.01062}
  (\bibinfo{year}{2020}).
\newblock


\bibitem[\protect\citeauthoryear{Wang, Xu, Gu, Song, and Green}{Wang
  et~al\mbox{.}}{2021}]%
        {r38}
\bibfield{author}{\bibinfo{person}{Jianhong Wang}, \bibinfo{person}{Wangkun
  Xu}, \bibinfo{person}{Yunjie Gu}, \bibinfo{person}{Wenbin Song}, {and}
  \bibinfo{person}{Tim~C. Green}.} \bibinfo{year}{2021}\natexlab{}.
\newblock \showarticletitle{Multi-Agent Reinforcement Learning for Active
  Voltage Control on Power Distribution Networks}.
\newblock \bibinfo{journal}{\emph{ArXiv}}  \bibinfo{volume}{abs/2110.14300}
  (\bibinfo{year}{2021}).
\newblock


\bibitem[\protect\citeauthoryear{Wang, Ye, and Lu}{Wang et~al\mbox{.}}{2022}]%
        {r39}
\bibfield{author}{\bibinfo{person}{Jiangxing Wang}, \bibinfo{person}{Deheng
  Ye}, {and} \bibinfo{person}{Zongqing Lu}.} \bibinfo{year}{2022}\natexlab{}.
\newblock \showarticletitle{More Centralized Training, Still Decentralized
  Execution: Multi-Agent Conditional Policy Factorization}.
\newblock \bibinfo{journal}{\emph{ArXiv}}  \bibinfo{volume}{abs/2209.12681}
  (\bibinfo{year}{2022}).
\newblock


\bibitem[\protect\citeauthoryear{Wang, Wang, Zheng, and Zhang}{Wang
  et~al\mbox{.}}{2019a}]%
        {r40}
\bibfield{author}{\bibinfo{person}{Tonghan Wang}, \bibinfo{person}{Jianhao
  Wang}, \bibinfo{person}{Chongyi Zheng}, {and} \bibinfo{person}{Chongjie
  Zhang}.} \bibinfo{year}{2019}\natexlab{a}.
\newblock \showarticletitle{Learning Nearly Decomposable Value Functions Via
  Communication Minimization}.
\newblock \bibinfo{journal}{\emph{ArXiv}}  \bibinfo{volume}{abs/1910.05366}
  (\bibinfo{year}{2019}).
\newblock


\bibitem[\protect\citeauthoryear{Wang, Wang, Zheng, and Zhang}{Wang
  et~al\mbox{.}}{2019b}]%
        {NDQ}
\bibfield{author}{\bibinfo{person}{Tonghan Wang}, \bibinfo{person}{Jianhao
  Wang}, \bibinfo{person}{Chongyi Zheng}, {and} \bibinfo{person}{Chongjie
  Zhang}.} \bibinfo{year}{2019}\natexlab{b}.
\newblock \showarticletitle{Learning nearly decomposable value functions via
  communication minimization}.
\newblock \bibinfo{journal}{\emph{arXiv preprint arXiv:1910.05366}}
  (\bibinfo{year}{2019}).
\newblock


\bibitem[\protect\citeauthoryear{Xu, Zhang, Li, Zhang, Zhou, and Fan}{Xu
  et~al\mbox{.}}{2022}]%
        {r41}
\bibfield{author}{\bibinfo{person}{Zhiwei Xu}, \bibinfo{person}{Bin Zhang},
  \bibinfo{person}{Dapeng Li}, \bibinfo{person}{Zeren Zhang},
  \bibinfo{person}{Guangchong Zhou}, {and} \bibinfo{person}{Guoliang Fan}.}
  \bibinfo{year}{2022}\natexlab{}.
\newblock \showarticletitle{Consensus Learning for Cooperative Multi-Agent
  Reinforcement Learning}. In \bibinfo{booktitle}{\emph{AAAI Conference on
  Artificial Intelligence}}.
\newblock


\bibitem[\protect\citeauthoryear{Zhang, Li, Wang, Xie, and Lu}{Zhang
  et~al\mbox{.}}{2021}]%
        {r42}
\bibfield{author}{\bibinfo{person}{Tianhao Zhang}, \bibinfo{person}{Yueheng
  Li}, \bibinfo{person}{Chen Wang}, \bibinfo{person}{Guangming Xie}, {and}
  \bibinfo{person}{Zongqing Lu}.} \bibinfo{year}{2021}\natexlab{}.
\newblock \showarticletitle{FOP: Factorizing Optimal Joint Policy of
  Maximum-Entropy Multi-Agent Reinforcement Learning}. In
  \bibinfo{booktitle}{\emph{International Conference on Machine Learning}}.
\newblock


\bibitem[\protect\citeauthoryear{Zhou, Luo, Villela, Yang, Rusu, Miao, Zhang,
  Alban, Fadakar, Chen, Huang, Wen, Hassanzadeh, Graves, Chen, Zhu, Nguyen,
  Elsayed, Shao, Ahilan, Zhang, Wu, Fu, Rezaee, Yadmellat, Rohani, Nieves, Ni,
  Banijamali, Rivers, Tian, Palenicek, Ammar, Zhang, Liu, Hao, and Wang}{Zhou
  et~al\mbox{.}}{2020}]%
        {r43}
\bibfield{author}{\bibinfo{person}{Ming Zhou}, \bibinfo{person}{Jun Luo},
  \bibinfo{person}{Julian Villela}, \bibinfo{person}{Yaodong Yang},
  \bibinfo{person}{David Rusu}, \bibinfo{person}{Jiayu Miao},
  \bibinfo{person}{Weinan Zhang}, \bibinfo{person}{Montgomery Alban},
  \bibinfo{person}{Iman Fadakar}, \bibinfo{person}{Zheng Chen},
  \bibinfo{person}{Aurora~Chongxi Huang}, \bibinfo{person}{Ying Wen},
  \bibinfo{person}{Kimia Hassanzadeh}, \bibinfo{person}{Daniel Graves},
  \bibinfo{person}{Dong Chen}, \bibinfo{person}{Zhengbang Zhu},
  \bibinfo{person}{Nhat~M. Nguyen}, \bibinfo{person}{Mohamed Elsayed},
  \bibinfo{person}{Kun Shao}, \bibinfo{person}{Sanjeevan Ahilan},
  \bibinfo{person}{Baokuan Zhang}, \bibinfo{person}{Jiannan Wu},
  \bibinfo{person}{Zhengang Fu}, \bibinfo{person}{Kasra Rezaee},
  \bibinfo{person}{Peyman Yadmellat}, \bibinfo{person}{Mohsen Rohani},
  \bibinfo{person}{Nicolas~Perez Nieves}, \bibinfo{person}{Yihan Ni},
  \bibinfo{person}{Seyedershad Banijamali}, \bibinfo{person}{Alexander~Cowen
  Rivers}, \bibinfo{person}{Zheng Tian}, \bibinfo{person}{Daniel Palenicek},
  \bibinfo{person}{Haitham Ammar}, \bibinfo{person}{Hongbo Zhang},
  \bibinfo{person}{Wulong Liu}, \bibinfo{person}{Jianye Hao}, {and}
  \bibinfo{person}{Jun Wang}.} \bibinfo{year}{2020}\natexlab{}.
\newblock \showarticletitle{SMARTS: Scalable Multi-Agent Reinforcement Learning
  Training School for Autonomous Driving}.
\newblock \bibinfo{journal}{\emph{ArXiv}}  \bibinfo{volume}{abs/2010.09776}
  (\bibinfo{year}{2020}).
\newblock


\bibitem[\protect\citeauthoryear{Zhou, Liu, Qing, Chen, Zheng, Huang, Song, and
  Song}{Zhou et~al\mbox{.}}{2023}]%
        {r44}
\bibfield{author}{\bibinfo{person}{Yihe Zhou}, \bibinfo{person}{Shunyu Liu},
  \bibinfo{person}{Yunpeng Qing}, \bibinfo{person}{Kaixuan Chen},
  \bibinfo{person}{Tongya Zheng}, \bibinfo{person}{Yanhao Huang},
  \bibinfo{person}{Jie Song}, {and} \bibinfo{person}{Mingli Song}.}
  \bibinfo{year}{2023}\natexlab{}.
\newblock \showarticletitle{Is Centralized Training with Decentralized
  Execution Framework Centralized Enough for MARL?}
\newblock \bibinfo{journal}{\emph{ArXiv}}  \bibinfo{volume}{abs/2305.17352}
  (\bibinfo{year}{2023}).
\newblock


\end{thebibliography}


\end{document}